\documentclass[conference]{IEEEtran}

\IEEEsettopmargin{t}{0.755in}

\usepackage{cite}
\usepackage{graphicx}
\usepackage{subcaption}
\usepackage{xcolor}
\usepackage{pgfplots}
\usepackage{multirow}
\usepackage{upgreek}
\usepackage{amssymb}
\usepackage{amsmath}
\usepackage{cases}
\usepackage{pifont}
\usepackage{bm}
\usepackage{amsthm}

\usepackage{tabularx}
\usepackage{booktabs}
\newcolumntype{Y}{>{\centering\arraybackslash}X}
\usepackage{array}
\usepackage{float}
\usepackage{xcolor}
\usepackage{algorithmic}
\usepackage[linesnumbered,ruled,vlined]{algorithm2e}
\usepackage{changes}
\usepackage{bbm}
\usepackage{soul}
\usetikzlibrary{arrows,positioning,shapes.geometric,spy}

\newcommand{\fixme}[2]{\ifx&#2&{\leavevmode\color{red}#1}\else{\leavevmode\color{red}FIXME\{}#1{\leavevmode\color{red}\}}\footnote{{\leavevmode\color{red}#2}}\PackageWarning{Fixme}{#1: #2}\fi}

\SetCommentSty{mycommfont}

\DeclareMathOperator*{\argmin}{arg\,min}
\DeclareMathOperator*{\argmax}{arg\,max}
\DeclareMathOperator*{\sgn}{sgn}

\DeclareMathOperator*{\relu}{ReLU}

\begin{document}

\title{Neural Dynamic Successive Cancellation \\ Flip Decoding of Polar Codes}

\author{
\IEEEauthorblockN{Nghia Doan\IEEEauthorrefmark{1}, Seyyed Ali Hashemi\IEEEauthorrefmark{2}, Furkan Ercan\IEEEauthorrefmark{1}, Thibaud Tonnellier\IEEEauthorrefmark{1}, Warren J. Gross\IEEEauthorrefmark{1}}
\IEEEauthorblockA{\IEEEauthorrefmark{1}Department of Electrical and Computer Engineering, McGill University, Canada} 
\IEEEauthorblockA{\IEEEauthorrefmark{2}Department of Electrical Engineering, Stanford University, USA}
\IEEEauthorblockA{\IEEEauthorrefmark{1}\{nghia.doan, furkan.ercan\}@mail.mcgill.ca, \{thibaud.tonnellier, warren.gross\}@mcgill.ca}
\IEEEauthorblockA{\IEEEauthorrefmark{2}ahashemi@stanford.edu}
}

\maketitle
\begin{abstract}
Dynamic successive cancellation flip (DSCF) decoding of polar codes is a powerful algorithm that can achieve the error correction performance of successive cancellation list (SCL) decoding, with a complexity that is close to that of successive cancellation (SC) decoding at practical signal-to-noise ratio (SNR) regimes. However, DSCF decoding requires costly transcendental computations which adversely affect its implementation complexity. In this paper, we first show that a direct application of common approximation schemes on the conventional DSCF decoding results in significant error-correction performance loss. We then introduce a training parameter and propose an approximation scheme which completely removes the need to perform transcendental computations in DSCF decoding, with almost no error-correction performance degradation.
\end{abstract}

\IEEEpeerreviewmaketitle
\section{Introduction} \label{sec:intro}
Polar codes are proven to achieve channel capacity for any binary symmetric channel under the low-complexity successive cancellation (SC) decoding  as the code length tends towards infinity \cite{arikan}. Recently, polar codes are selected for use in the enhanced mobile broadband (eMBB) control channel of the fifth generation of cellular mobile communications (5G standard) which requires codes of short length \cite{3gpp_report}. The error-correction performance of SC decoding for short polar codes does not satisfy the requirements of the 5G standard. A SC list (SCL) decoding was introduced in \cite{tal_list} to improve the error-correction performance of SC decoding for short to moderate polar codes by keeping a list of candidate codewords at each decoding step. In addition, it was observed that under SCL decoding, the error probabilities are significantly enhanced when the polar code is concatenated with a cyclic redundancy check (CRC) \cite{tal_list}. However, the implementation complexity of SCL decoding grows as the list size increases.

SC flip (SCF) decoding algorithm was introduced in \cite{SCF} which unlike SCL decoding, performs multiple SC decoding attempts in series, where each decoding attempt tries to flip the first-order erroneous information bit of the previous decoding attempt. Similar to SCL decoding, SCF decoding relies on a CRC to indicate whether the decoding is successful or not. Several methods have been recently proposed to improve the error-correction performance of SCF \cite{PSCF-ICC18,Carlo_SCFlip,SCFlip_TCOM18}, however they are limited with correcting a single erroneous bit in the codeword. Dynamic SCF (DSCF) decoding \cite{DSCF} is a generalization of SCF-based decoding which is able to correct higher-order erroneous information bits, i.e., DSCF decoding can correct an erroneous bit which is a result of error propagation under SC decoding, given that all previous erroneous bits were correctly flipped \cite{DSCF}.

The advantage of DSCF decoding is that the average decoding complexity of it at high signal-to-noise ratio (SNR) regimes asymptotically approaches the decoding complexity of SC decoding, while maintaining an error-correction performance \cite{DSCF} comparable to that of SCL decoding. However, DSCF requires costly exponential and logarithmic computations which prevent the algorithm to be attractive for practical applications.

In this paper, we first show that a direct application of the common approach to approximate the underlying exponential and logarithmic function in the DSCF decoding algorithm results in a significant error-correction performance degradation. We then introduce a new trainable perturbation parameter to the DSCF decoding algorithm and show that the resulting decoder can have an error-correction performance close to that of the ideal DSCF decoder. Furthermore, we show that the proposed decoder does not suffer from significant error-correction performance loss if the common hardware-friendly approximation techniques are used. In addition, a novel deep learning framework which utilizes the symmetric properties of DSCF decoding is introduced as an optimization scheme for the trainable parameter. We name the proposed decoding algorithm neural DSCF (NDSCF) decoding algorithm. Simulation results show that for a 5G polar code of length $256$, with $128$ information bits and concatenated with a $24$-bit CRC, the proposed NDSCF decoding does not incur significant error correction performance loss in comparison with the ideal DSCF decoding, while requiring no exponential or logarithmic computations.

The rest of this paper is organized as follows. Section~\ref{sec:polar} briefly reviews polar codes and DSCF decoding. Section~\ref{sec:NDSCF} describes the proposed NDSCF decoder. Section~\ref{sec:experiment} provides numerical results, and finally, Section~\ref{sec:conclude} presents concluding remarks.

\section{Preliminaries}
\label{sec:polar}

\subsection{Polar Codes}
\label{sec:polar:polar}
A polar code $\mathcal{P}(N,K)$ of length $N$ with $K$ information bits is constructed by applying a linear transformation to the message word $\bm{u} = \{u_0,u_1,\ldots,u_{N-1}\}$ as $\bm{x} = \bm{u}\bm{G}^{\otimes n}$, where $\bm{x} = \{x_0,x_1,\ldots,x_{N-1}\}$ is the codeword, $\bm{G}^{\otimes n}$ is the $n$-th Kronecker power of the polarizing matrix $\bm{G}=\bigl[\begin{smallmatrix} 1&0\\ 1&1 \end{smallmatrix} \bigr]$, and $n = \log_2 N$. The vector $\bm{u}$ contains a set $\mathcal{A}$ of $K$ information bits and a set $\mathcal{A}^c$ of $N-K$ frozen bits. The positions of the frozen bits are known to the encoder and the decoder and their values are usually set to $0$. The codeword $\bm{x}$ is then sent through the channel using binary phase-shift keying (BPSK) modulation. The soft vector of the transmitted codeword received by the decoder is $\bm{y}=(\mathbf{1}-2\bm{x})+\bm{z}$, where $\mathbf{1}$ is an all-one vector of size $N$, and $\bm{z} \in \mathbb{R}^N$ is the additive white Gaussian noise (AWGN) vector with variance $\sigma^2$ and zero mean. In the log-likelihood ratio (LLR) domain, the LLR vector of the transmitted codeword is $\bm{L}_{n} = \frac{2\bm{y}}{\sigma^2}$.

\subsection{Successive Cancellation Decoding}	
\label{sec:polar:SC}

\begin{figure}[t]
	\centering
	\begin{subfigure}[b]{0.5\textwidth}
		\centering
		\includegraphics[width=0.6\linewidth]{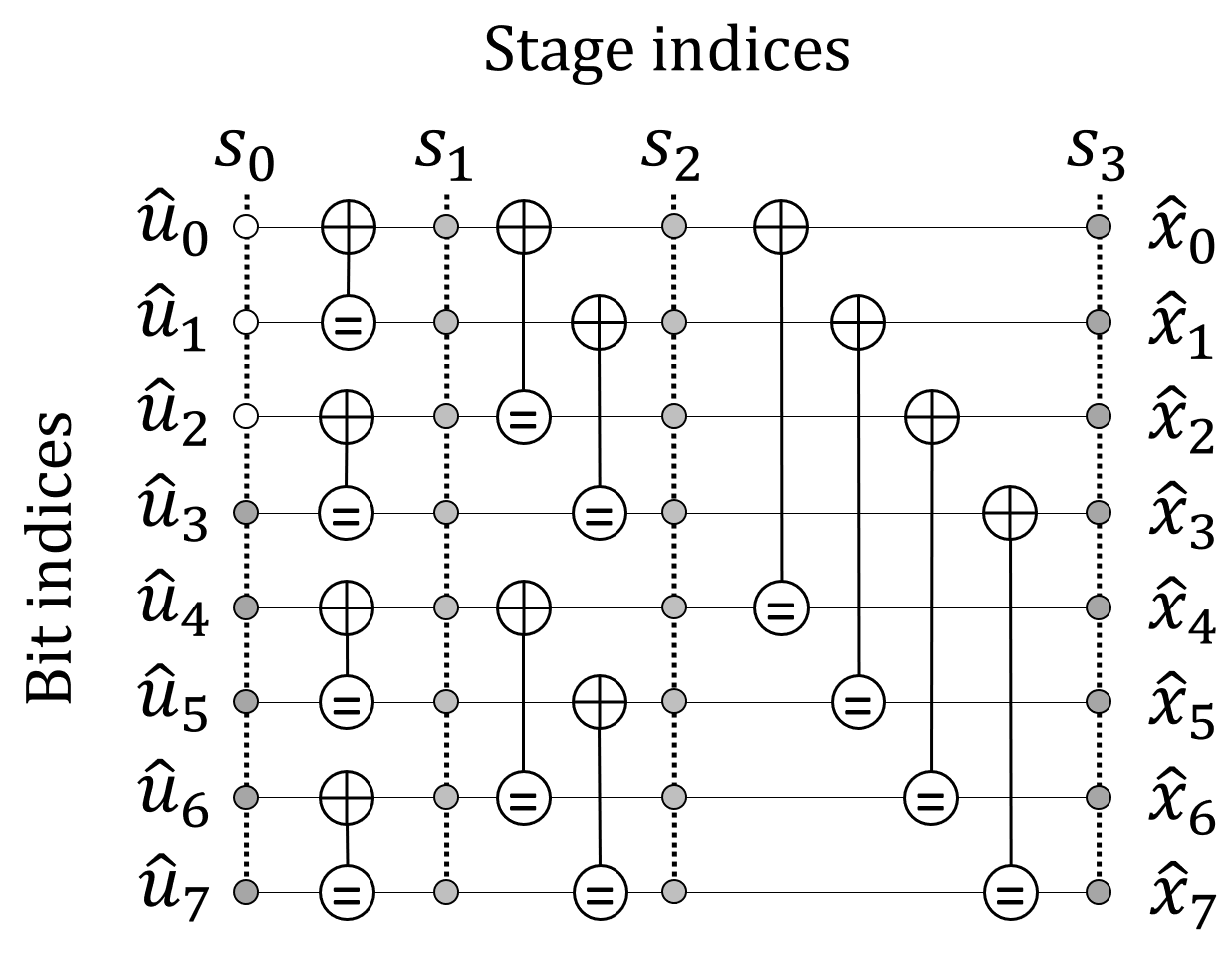}
		\caption{}
		\vspace*{5pt}
		\label{fig:SCGraph}
	\end{subfigure}
	\begin{subfigure}[b]{0.5\textwidth}			
		\centering
		\hspace*{15pt}
		\includegraphics[width=0.6\linewidth]{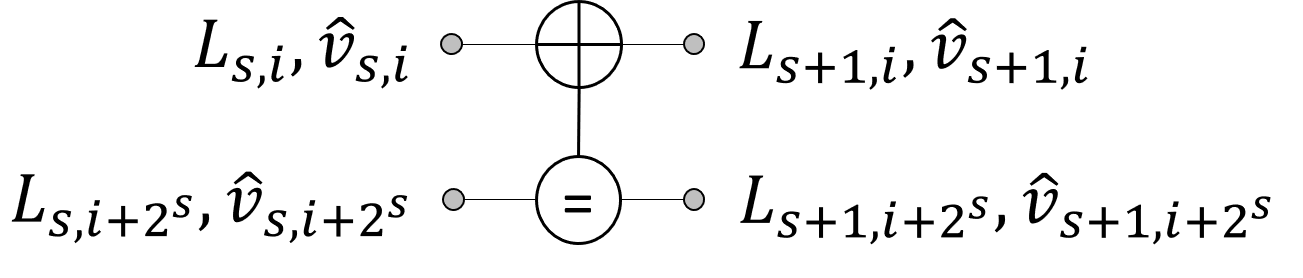}
		\caption{}
		\label{fig:SCPE}
	\end{subfigure}  
	\caption{(a) SC decoding on the factor graph of $\mathcal{P}(8,5)$ with $\{u_0,u_1,u_2\}\in \mathcal{A}^c$, (b) a PE.}
	\vspace*{-1\baselineskip}
\end{figure}

SC decoding can be illustrated on a polar code factor graph representation. An example of a factor graph for $\mathcal{P}(8,5)$ is depicted in Fig.~\ref{fig:SCGraph}. To obtain the message word, the soft LLR values and the hard bit estimations are propagated through all the processing elements (PEs), which are depicted in Fig.~\ref{fig:SCPE}. The computation of a PE is obtained as
\begin{equation}
\label{equ:SCPE_L}
\begin{cases}
L_{s,i} = f(L_{s+1,i},L_{s+1,i+2^s})\\
L_{s,i+2^s} = g(L_{s+1,i},L_{s+1,i+2^s},\hat{v}_{s,i})
\end{cases}
\end{equation}
where 
\begin{equation*}
\begin{cases}
f(a,b) = \min(|a|,|b|)\sgn(a)\sgn(b), \\
g(a,b,c) = b + (1-2c)a, \\
\end{cases}
\end{equation*}
and $L_{s,i}$ and $\hat{v}_{s,i}$ are the soft LLR value and the hard-bit estimation at the $s$-th stage and the $i$-th bit, respectively. The hard-bit values of the PE are computed as
\begin{equation}
\label{equ:SCPE_v}
\begin{cases}
\hat{v}_{s+1,i} = \hat{v}_{s,i} \oplus \hat{v}_{s,i+2^s} \\
\hat{v}_{s+1,i+2^s} = \hat{v}_{s,i+2^s}.
\end{cases}
\end{equation}
The soft LLR values at the $n$-th stage are initialized to $\bm{L}_n$ and the hard-bit estimation at the $0$-th stage is obtained as
\begin{equation}
\label{equ:SC:HardDecision}
\hat{u}_{i} = \hat{v}_{0,i}=
\begin{cases}
0 & \text{if } u_i \in \mathcal{A}^c,\\
\frac{1 - \sgn(L_{0,i})}{2} & \text{otherwise.}
\end{cases}
\end{equation}

\subsection{Dynamic Successive Cancellation Flip Decoding}

\label{sec:polar:DSCF}

The error-correction performance of SC decoding for short to moderate block lengths is not satisfactory. In order to improve its error-correction performance, a CRC of length $c$ is concatenated to the message word of polar codes to check whether SC decoding succeeded or not. If the estimated message word $\bm{\hat{u}}$ does not satisfy the CRC after the initial SC decoding attempt, a secondary SC decoding attempt is made by flipping the estimation of an information bit in $\bm{\hat{u}}$ which is most likely to be erroneous. This process can be performed multiple times by applying a predetermined number of SC decoding attempts with each attempt flipping the estimation of a different information bit. If the resulting message word after one of the SC decoding attempts satisfies the CRC, the decoding is declared successful. This algorithm is referred to as SCF decoding \cite{SCF}. The main problem associated with SCF decoding is that only the first error bit after the initial SC decoding can be corrected. However, it is common that even after the first error bit is corrected, the resulting message word still contains error bits. Therefore, further flipping attempts for the additional error bits are required. DSCF decoding was introduced in \cite{DSCF} to address this problem.

Let $\mathcal{E}_\omega = \{i_1,\dots,i_\omega\}$, where $\{u_{i_1},\dots,u_{i_\omega}\} \subset \mathcal{A}$, be the set of bit-flipping positions of order $\omega$, where $i_1 < \dots < i_\omega$, $0 \leq \omega \leq K+c$, and $|\mathcal{E}_\omega| = \omega$. Note that $\mathcal{E}_0 = \emptyset$. In the course of DSCF decoding, the hard-bit estimations of all the bit indices in $\mathcal{E}_\omega$ are flipped which can be written as
\begin{equation}
\label{equ:DSCF:HardDecision}
\hat{u}[{\mathcal{E}_{\omega-1}}]_{i} =
\begin{cases}
0 & \text{if } u_i \in \mathcal{A}^c,\\
\frac{1 + \sgn(L_{0,i})}{2} & \text{if $u_i \in \mathcal{A}, i \in \mathcal{E}_\omega$,}\\
\frac{1 - \sgn(L_{0,i})}{2} & \text{otherwise.} \\
\end{cases}
\end{equation}
The set $\mathcal{E}_\omega$ is constructed progressively based on the set $\mathcal{E}_{\omega-1} = \{i_1,\dots,i_{\omega-1}\}$. In fact, if SC decoding fails after flipping all the bit-flipping positions in $\mathcal{E}_{\omega-1}$, $i_\omega$ is added to $\mathcal{E}_{\omega-1}$ to form $\mathcal{E}_\omega$ and an additional SC decoding attempt is performed by flipping the bit estimation at all the bit-flipping positions in $\mathcal{E}_\omega$. Furthermore, a maximum number of decoding attempts $m_\omega$ is imposed on the decoder to limit the computational complexity in practice.

Let
\begin{equation}
\label{equ:DSCF:p_star}
p^*_i({\mathcal{E}_{\omega-1}}) = \text{Pr}(\hat{u}[{\mathcal{E}_{\omega-1}}]_i=u_i| \bm{y},  \bm{\hat{u}}[{\mathcal{E}_{\omega-1}}]_0^{i-1} = \bm{u}_0^{i-1}),
\end{equation}
where $\bm{\hat{u}}[{\mathcal{E}_{\omega-1}}]_0^{i-1} = \{\hat{u}[\mathcal{E}_{\omega-1}]_{0}, \hat{u}[\mathcal{E}_{\omega-1}]_{1}, \dots, \hat{u}[\mathcal{E}_{\omega-1}]_{i-1}\}$ and $\bm{u}_0^{i-1} = \{u_0,u_1,\dots,u_{i-1}\}$. The probability that SC decoding is successful after flipping all the bit-flipping positions in $\mathcal{E}_\omega$ is then defined as \cite{DSCF}
\begin{equation}
\label{equ:DSCF:P_flip}
P^*(\mathcal{E}_\omega)=\prod_{\substack{{\forall i \in \mathcal{A}\setminus\mathcal{E}_\omega}\\
i < i_\omega}} p^*_i({\mathcal{E}_{\omega-1}}) \times
\prod_{\forall i \in \mathcal{E}_\omega} \left( 1 - p^*_i({\mathcal{E}_{\omega-1}}) \right).
\end{equation}
Therefore, the bit-flipping position $i^*_\omega$ that maximizes the probability of $\bm{\hat{u}}[{\mathcal{E}_{\omega-1}}]$ being correctly decoded is
\begin{equation}
\label{equa:DSCF:t_flip}
i^*_\omega = \argmax_{\substack{\forall i_\omega, i_{\omega-1} < i_\omega \leq N-1, u_{i_\omega} \in \mathcal{A} \\ \mathcal{E}_\omega = \mathcal{E}_{\omega-1} \cup i_\omega }} P^*(\mathcal{E}_\omega).
\end{equation}
Note that the probability $p^*_i({\mathcal{E}_{\omega-1}})$ cannot be obtained during the course of decoding as the values of the elements of $\bm{u}$ are unknown to the decoder \cite{DSCF}. As a result, DSCF uses a known probability $p_i({\mathcal{E}_{\omega-1}})$ to estimate $p^*_i({\mathcal{E}_{\omega-1}})$, which is defined as
\begin{equation}
\label{equ:DSCF:p_estimate}
\begin{split}
p_i({\mathcal{E}_{\omega-1}}) 
&= \max \big(\text{Pr}(\hat{u}[{\mathcal{E}_{\omega-1}}]_i=0| \bm{y},\bm{\hat{u}}[{\mathcal{E}_{\omega-1}}]_0^{i-1}),\\
& \hspace*{38pt} \text{Pr}(\hat{u}[{\mathcal{E}_{\omega-1}}]_i=1| \bm{y}, \bm{\hat{u}}[{\mathcal{E}_{\omega-1}}]_0^{i-1})\big) \\
&= \frac{1}{1+\exp\left(-|L[\mathcal{E}_{\omega-1}]_{0,i}|\right)},\\
\end{split}
\end{equation}
where $L[\mathcal{E}_{\omega-1}]_{0,i}$ is the corresponding LLR value of $\hat{u}[\mathcal{E}_{\omega-1}]_{i}$. It was shown in \cite{DSCF} that the estimation in (\ref{equ:DSCF:p_estimate}) is not accurate. Therefore, \cite{DSCF} introduced a perturbation parameter $\alpha$ to have a better estimation of $p^*_i({\mathcal{E}_{\omega-1}})$ as
\begin{equation}
\label{equ:DSCF:p_star_estimate}
p^*_i({\mathcal{E}_{\omega-1}}) \approx \frac{1}{1+\exp\left(-\alpha|L[\mathcal{E}_{\omega-1}]_{0,i}|\right)}\text{.}
\end{equation}
It should be noted that $\alpha \in \mathbb{R}^+$ is a scaling factor for the magnitude of the LLR values and is determined by a Monte-Carlo simulation. To enable a trade-off between decoding latency and error-correction performance, instead of only flipping the most probable bit-flipping position, DSCF decoding attempts to improve SC decoding with a list of most probable bit-flipping indices $i^*_\omega$ at each error order $\omega$ \cite{DSCF}.

In order to have numerically stable computations in the hardware implementation of the DSCF decoder, the bit-flipping metric in (\ref{equ:DSCF:P_flip}) can be written in the log-likelihood (LL) domain as
\begin{equation}
\label{equ:DSCF:DSCF_LL}
\begin{split}
Q^*(\mathcal{E}_\omega) & = -\frac{1}{\alpha}\ln(P^*(\mathcal{E}_\omega)) \\
& = \sum_{\substack{{\forall i \in \mathcal{A}}\\ i \leq i_\omega}} \frac{1}{\alpha}\ln{\left(1+\exp\left(-\alpha|L[\mathcal{E}_{\omega-1}]_{0,i}|\right)\right)} \\
& \hspace*{12pt} + \sum_{\forall i \in \mathcal{E}_\omega} |L[\mathcal{E}_{\omega-1}]_{0,i}|.
\end{split}
\end{equation}
Consequently, the most probable bit-flipping position $i^*_\omega$ can be found in the LL domain as
\begin{equation}
\label{equ:NDSCF:sel}
i^*_\omega = \argmin_{\substack{\forall i_\omega, i_{\omega-1} < i_\omega \leq N-1, u_{i_\omega} \in \mathcal{A} \\ \mathcal{E}_\omega = \mathcal{E}_{\omega-1} \cup i_\omega }} Q^*(\mathcal{E}_{\omega}).
\vspace{1pt}
\end{equation}

\section{Neural Dynamic Successive Cancellation Flip Decoding}
\label{sec:NDSCF}

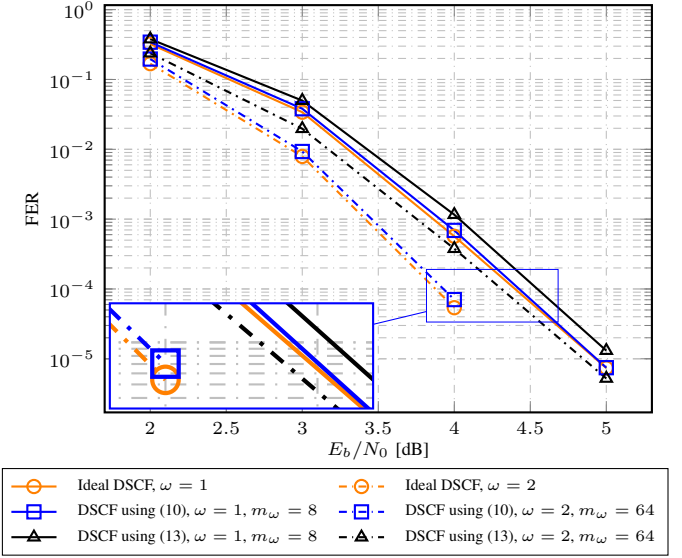
\begin{figure}[t]
	\centering
	\begin{tikzpicture}[spy using outlines = {rectangle, magnification=2.0, connect spies}]
\pgfplotsset{	
	label style = {font=\fontsize{7pt}{7}\selectfont},
	tick label style = {font=\fontsize{7pt}{7}\selectfont}
}

\begin{axis}[
scale = 1,
ymode=log,
xlabel={$E_b/N_0$ [\text{dB}]}, xlabel style={yshift=0.8em},
ylabel={FER}, ylabel style={yshift=-0.75em},
xtick={1,1.5,2,2.5,3,3.5,4,4.5,5},
ytick={1e-5,1e-4,1e-3,1e-2,1e-1,1e0},
grid=both,
ymajorgrids=true,
xmajorgrids=true,
grid style=dashdotted,
width=1\columnwidth, height=7cm,
thick,
mark size=2.5,
legend cell align={left},
legend style={
	at={(0,1e-5)},
	anchor=south west,
	column sep= 2mm,
	font=\fontsize{6pt}{7.2}\selectfont,
},
  legend to name=perf-legend-DSCF,
  legend columns=2,
]

\addplot[
color=orange,
solid,
mark=o,
thick,
mark size=2.5,
]
table {
	2	0.322
	3	0.034
	4	0.000567568
	5	7.38E-06
};
\addlegendentry{Ideal DSCF, $\omega=1$}

\addplot[
color=orange,
dashdotted,
mark=o,
mark options={solid},
thick,
mark size=2.5,
]
table {
	2	0.1683
	3	0.0079
	4	5.39E-05
};
\addlegendentry{Ideal DSCF, $\omega=2$}

\addplot[
color=blue,
solid,
mark=square,
thick,
mark size=2.5,
]
table {
	2	0.3422
	3	0.0381
	4	0.000689655
	5	7.46E-06
};
\addlegendentry{DSCF using (\ref{equ:DSCF:DSCF_LL}), $\omega=1$, $m_\omega=8$}

\addplot[
color=blue,
dashdotted,
mark=square,
mark options={solid},
thick,
mark size=2.5,
]
table {
	2	0.1962
	3	9.30E-03
	4	7.05E-05
};
\addlegendentry{DSCF using (\ref{equ:DSCF:DSCF_LL}), $\omega=2$, $m_\omega=64$}

\addplot[
color=black,
solid,
mark=triangle,
thick,
mark size=2.5,
]
table {	
	2	0.3789
	3	0.0498
	4	0.001157895
	5	1.31E-05	
};
\addlegendentry{DSCF using (\ref{equ:DSCF:DSCF_LL_HW}), $\omega=1$, $m_\omega=8$}

\addplot[
color=black,
dashdotted,
mark=triangle,
mark options={solid},
thick,
mark size=2.5,
]
table {
	2	0.2381
	3	1.97E-02
	4	0.000377358	
	5	5.23E-06
};
\addlegendentry{DSCF using (\ref{equ:DSCF:DSCF_LL_HW}), $\omega=2$, $m_\omega=64$}

\coordinate (spypoint1) at (axis cs:4.25,0.8e-4);
\coordinate (magnifyglass1) at (axis cs:2.6,1.1e-5);

\coordinate (spypoint2) at (axis cs:4,0.7e-3);
\coordinate (magnifyglass2) at (axis cs:4.8,7e-2);
\end{axis}
\spy [blue, width=3.5cm, height=1.4cm] on (spypoint1) in node[fill=white] at (magnifyglass1);

\end{tikzpicture}
	\ref{perf-legend-DSCF}
	\caption{Effect of the simplification in (\ref{equ:DSCF:DSCF_LL_HW}) on the FER of DSCF decoding for $\mathcal{P}(256,128)$, concatenated with a $24$-bit CRC. The ideal DSCF decoder is also plotted as a reference.}
	\label{fig:DSCF:FER}
	\vspace*{-1\baselineskip}
\end{figure}

Efficient hardware implementation of DSCF decoding is based on efficient implementation of the bit-flipping metric in (\ref{equ:DSCF:DSCF_LL}). However, (\ref{equ:DSCF:DSCF_LL}) involves logarithmic and exponential functions which are not hardware friendly. A common approach to approximate the logarithmic and exponential function in (\ref{equ:DSCF:DSCF_LL}) is to use the rectifier linear unit (ReLU) as \cite{Alexios_LLR_SCLD}
\begin{equation}
\ln(1+\exp(x)) \approx \relu(x) = 
\begin{cases}
x & \text{if $x>0$,} \\
0 & \text{otherwise.}\\
\end{cases}
\label{equ:relu}
\end{equation}
However, since $\alpha>0$, $-\alpha|L[\mathcal{E}_{\omega-1}]_{0,i}|<0$. Therefore, (\ref{equ:DSCF:DSCF_LL}) can be simplified as
\begin{align}
\label{equ:DSCF:DSCF_LL_HW}
Q^*(\mathcal{E}_\omega) \approx & \sum_{\substack{{\forall i \in \mathcal{A}}\\ i \leq i_\omega}} \frac{1}{\alpha}\relu\left(-\alpha|L[\mathcal{E}_{\omega-1}]_{0,i}|\right) \nonumber \\
&+ \sum_{\forall i \in \mathcal{E}_\omega} |L[\mathcal{E}_{\omega-1}]_{0,i}| \nonumber \\
= & \sum_{\forall i \in \mathcal{E}_\omega} |L[\mathcal{E}_{\omega-1}]_{0,i}|\text{,}
\end{align}
which is independent of the perturbation parameter $\alpha$. Fig.~\ref{fig:DSCF:FER} shows the effect of the simplification in (\ref{equ:DSCF:DSCF_LL_HW}) on the error-correction performance of DSCF decoding in terms of frame error rate (FER) for $\mathcal{P}(256,128)$ concatenated with a $24$-bit CRC which is used in 5G standard \cite{3gpp_report}. In this figure, $\alpha=0.3367$ and is calculated as described in \cite{DSCF}. The FER of the ideal DSCF decoder where the erroneous bits up to the $\omega$-th error order can always be accurately corrected is also plotted for comparison. It can be seen that using (\ref{equ:DSCF:DSCF_LL_HW}) incurs $0.1$~dB and $0.4$~dB of FER performance loss for DSCF decoding in comparison with using (\ref{equ:DSCF:DSCF_LL}), when $\omega=1$ and $\omega=2$ respectively, at a target FER of $10^{-4}$.

The main reason that results in such an error-correction performance degradation is that the perturbation parameter $\alpha$ is a multiplicative positive parameter which renders the ReLU function to be zero independent of the LLR value. To address this issue, we propose to use a perturbation parameter $\beta \in \mathbb{R}^+$ which unlike $\alpha$, is an additive positive parameter, and like $\alpha$, tries to improve the estimation of $p^*_i({\mathcal{E}_{\omega-1}})$. We write the proposed estimation of $p^*_i({\mathcal{E}_{\omega-1}})$ as
\begin{equation}
\label{equ:NDSCF:p_star_estimate}
p^*_i({\mathcal{E}_{\omega-1}}) \approx \frac{1}{1+\exp\left(\beta-|L[\mathcal{E}_{\omega-1}]_{0,i}|\right)}.
\end{equation}
In addition, we propose to use a bit-flipping metric in the LL domain $Q^*(\mathcal{E}_\omega)$ which is tailored to the proposed $p^*_i({\mathcal{E}_{\omega-1}})$ in (\ref{equ:NDSCF:p_star_estimate}) as
\begin{equation}
\label{equ:NDSCF:NDSCF_LL}
\begin{split}
Q^*(\mathcal{E}_\omega) & = -\ln(P^*(\mathcal{E}_\omega)) + \omega\beta \\
& = \sum_{\substack{{\forall i \in \mathcal{A}}\\ i \leq i_\omega}} \ln{\left(1+\exp\left(\beta - |L[\mathcal{E}_{\omega-1}]_{0,i}|\right)\right)} \\
& \hspace*{12pt} + \sum_{\forall i \in \mathcal{E}_\omega} |L[\mathcal{E}_{\omega-1}]_{0,i}|\text{,}
\end{split}
\end{equation}
where we used the fact that $\omega\beta$ is a constant and it will not affect the selection of $i^*_\omega$ in (\ref{equ:NDSCF:sel}).

Let us now use the ReLU function in (\ref{equ:relu}) to simplify the proposed bit-flipping metric in (\ref{equ:NDSCF:NDSCF_LL}) as
\begin{align}
\label{equ:NDSCF:NDSCF_HW_LL}
Q^*(\mathcal{E}_\omega) \approx & \sum_{\substack{{\forall i \in \mathcal{A}}\\ i \leq i_\omega}} \relu{\left(\beta - |L[\mathcal{E}_{\omega-1}]_{0,i}|\right)} \nonumber \\
& + \sum_{\forall i \in \mathcal{E}_\omega} |L[\mathcal{E}_{\omega-1}]_{0,i}| \nonumber \\
= & \sum_{\substack{{\forall i \in \mathcal{A}}\\ i \leq i_\omega \\ |L[\mathcal{E}_{\omega-1}]_{0,i}|<\beta}} \beta - |L[\mathcal{E}_{\omega-1}]_{0,i}| \nonumber \\
& + \sum_{\forall i \in \mathcal{E}_\omega} |L[\mathcal{E}_{\omega-1}]_{0,i}|\text{,}
\end{align}
where we used the fact that if $\beta-|L[\mathcal{E}_{\omega-1}]_{0,i}|>0$, then $\relu{\left(\beta - |L[\mathcal{E}_{\omega-1}]_{0,i}|\right)} = \beta - |L[\mathcal{E}_{\omega-1}]_{0,i}|$. It can be seen that the resulting hardware-friendly bit-flipping metric is dependent on the value of $\beta$.

We now show how to select the value of $\beta$. In conventional DSCF decoding of \cite{DSCF}, finding the optimal value of $\alpha$ in (\ref{equ:DSCF:DSCF_LL}) was treated as a search problem and was solved by a Monte-Carlo simulation. In this paper, we propose to treat the optimization of $\beta$ in (\ref{equ:NDSCF:NDSCF_LL}) or (\ref{equ:NDSCF:NDSCF_HW_LL}) as a learning problem and therefore, call the proposed algorithm NDSCF decoding. In order to do this, we use the inherent symmetric properties of the conventional DSCF decoding algorithm, which greatly simplify the training process as observed in \cite{Nachmani_STSP,Loren_ISIT_2017,Doan_ICC19}.

Note that since $p_i({\mathcal{E}_{\omega-1}})$ is independent from $\bm{u}$, the perturbed formulations of $p^*_i({\mathcal{E}_{\omega-1}})$ in (\ref{equ:DSCF:p_star_estimate}) and (\ref{equ:NDSCF:p_star_estimate}) are also independent from $\bm{u}$. Consequently, $P^*(\mathcal{E}_\omega)$ and $Q^*(\mathcal{E}_\omega)$ are also independent from $\bm{u}$. Therefore, all-zero codewords can be used for the training of $\beta$ in (\ref{equ:NDSCF:NDSCF_LL}) or (\ref{equ:NDSCF:NDSCF_HW_LL}). In addition, unlike the well-known classification problem in supervised learning \cite{DeepLearning}, obtaining the output labels for the correct bit-flipping positions $i^*_\omega$ is not mandatory in the proposed framework. This is because if the estimation of bit $i^*_\omega$ is flipped to have a correct value for the next SC decoding attempt, the message word given by the next SC decoding attempt will have a correct value at the position $i^*_\omega$, which is known to be $0$ since all-zero codewords are used. Instead of calculating an estimation error between $i^*_\omega$ and the true flipping position, we propose to use the output of SC decoding to calculate the objective loss function. Therefore, the heavy tasks of collecting the true flipping labels and representing them as one-hot encoded vectors are eliminated. As no labeling task is required for the error bits and only all-zero codewords are needed during the training, the proposed framework can be used in applications which require channel adaptation, where the decoder can be trained online to adapt with dynamic changes of the communication channel.

\begin{figure}[t]
	\centering
	\includegraphics[width=1\linewidth]{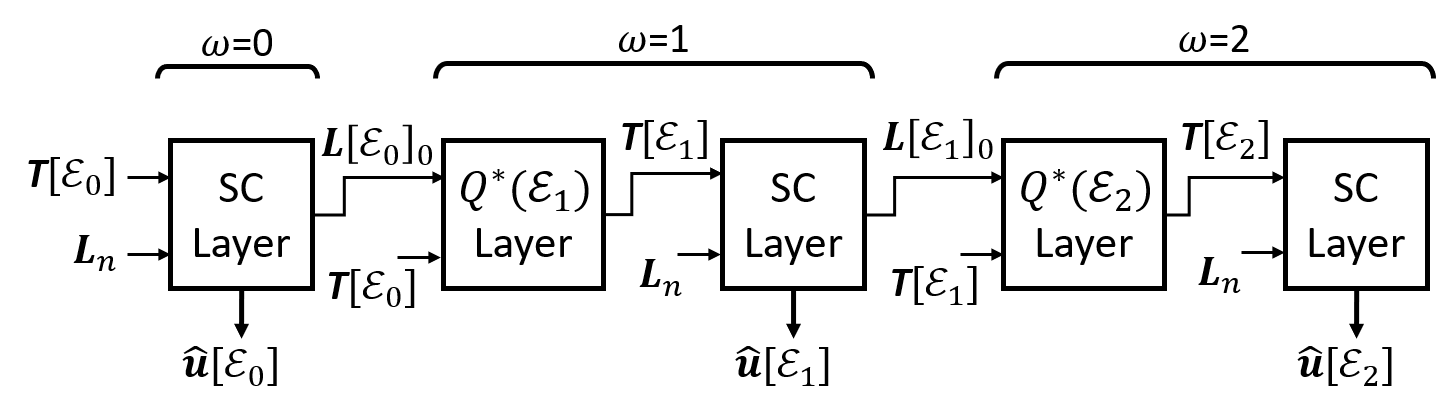}
	\caption{NDSCF decoder training framework with $\omega\in\{0,1,2\}$.}
	\label{fig:NDSCF}
	\vspace*{-1\baselineskip}
\end{figure}

Fig.~\ref{fig:NDSCF} illustrates the training framework of the proposed NDSCF decoder with $\omega\in\{0,1,2\}$. The framework contains two types of network layers, namely, SC and $Q^*(\mathcal{E}_\omega)$ layers. The SC layer performs standard SC decoding and the bit-flipping operations for all the bit indices in $\mathcal{E}_\omega$. On the other hand, the $Q^*(\mathcal{E}_\omega)$ layer is in charge of estimating the next bit-flipping position given that the estimated message word of the SC layer at error order $\omega-1$ does not satisfy the CRC. However, as we use a batch of channel LLR values $\bm{L}_n$ for training, the decoding of different $\bm{L}_n$ instances can be terminated at different error orders. This problem complicates the  implementation of the training framework with recent deep learning libraries \cite{Tensorflow}. To address this problem, we treat the decoding of all the $\bm{L}_n$ instances as the worst case scenario, where the correct message word can only be obtained after fixing the erroneous bit of the maximum considered error order. If a message word satisfies the CRC before reaching the last SC layer, the remaining $Q^*(\mathcal{E}_\omega)$ layers will maintain all the bit-flipping decisions of the successful CRC verification. Therefore, we define a bit-flipping vector $\bm{T}[\mathcal{E}_\omega] = \{T[\mathcal{E}_\omega]_0,\ldots,T[\mathcal{E}_\omega]_{N-1}\}$, which stores the bit-flipping decisions of all the information bits for the $\omega$-th SC layer, with $T[\mathcal{E}_\omega]_i \in \{-1,+1\}$ such that a value of $-1$ indicates a bit flip and a value of $+1$ indicates no bit flip.

The training framework starts with the initial SC decoding at $\omega=0$, which estimates the message word $\bm{\hat{u}}[\mathcal{E}_0]$ and the information bit LLR values $\bm{L}[\mathcal{E}_0]_0$. Note that $|\mathcal{E}_0|=\emptyset$ and $\bm{T}[\mathcal{E}_0] = \bm{1}$ where $\bm{1}$ is the all-ones vector of length $N$. If $\bm{\hat{u}}[\mathcal{E}_0]$ satisfies the CRC, the $Q^*(\mathcal{E}_1)$ and the $Q^*(\mathcal{E}_2)$ layers will not alter the bit-flipping vector and therefore set $\bm{T}[\mathcal{E}_1]=\bm{1}$ and $\bm{T}[\mathcal{E}_2]=\bm{1}$, respectively. As a result, $\bm{\hat{u}}[\mathcal{E}_1]$ and $\bm{\hat{u}}[\mathcal{E}_2]$ will have the same decoding values as $\bm{\hat{u}}[\mathcal{E}_0]$. On the other hand, if $\bm{\hat{u}}[\mathcal{E}_0]$ does not satisfy the CRC, given the input $\bm{L}[\mathcal{E}_0]_0$, the $Q^*(\mathcal{E}_1)$ layer computes either (\ref{equ:NDSCF:NDSCF_LL}) or (\ref{equ:NDSCF:NDSCF_HW_LL}) for all the bit indices $i_1$ of all the candidates, where $0 \leq i_1 \leq N-1$ and $u_{i_1} \in \mathcal{A}$. The $Q^*(\mathcal{E}_1)$ layer then selects $i^*_1$ based on (\ref{equ:NDSCF:sel}) and forms the bit-flipping vector $\bm{T}[\mathcal{E}_1]$ as
\begin{equation}
\label{equ:NDSCF:T_example}
T[\mathcal{E}_1]_i = 
\begin{cases}
+1 & \text{if } 0 \leq i \leq N-1 \text{, } i \neq i^*_1,\\
-1 & \text{if } i = i^*_1,
\end{cases}
\end{equation}
The bit-flipping vector $\bm{T}[\mathcal{E}_1]$ is then fed into the SC layer at error order $\omega=1$, which performs a standard SC decoding given the channel LLR values $\bm{L}_n$ and estimates $\bm{u}[\mathcal{E}_1]$ based on $\bm{T}[\mathcal{E}_1]$ as
\begin{equation}
\label{equ:NDSCF:u_example}
\hat{u}[\mathcal{E}_1]_{i} =
\begin{cases}
\frac{1 - \sgn(L[\mathcal{E}_1]_{0,i})T[\mathcal{E}_1]_i}{2} & \text{if } u[\mathcal{E}_1]_i \in \mathcal{A}\text{,} \\
0 & \text{otherwise.}
\end{cases}
\end{equation}
The decoding then continues in the same manner for all the layers at error order $\omega=2$. It is worth mentioning that the SC layer is implemented by unfolding the decoding process specified in (\ref{equ:SCPE_L}), (\ref{equ:SCPE_v}), and (\ref{equ:NDSCF:u_example}) at all the required stages and bit indices.

By induction, we can write general expressions for any error order $\omega$ for $\bm{T}[\mathcal{E}_\omega]$ as
\begin{equation}
\label{equ:NDSCF:T_generalized}
T[\mathcal{E}_\omega]_i = 
\begin{cases}
T[\mathcal{E}_{\omega-1}]_i & \text{if } 0 \leq i \leq i_{\omega-1},  \\
+1 & \text{if CRC$(\bm{\hat{u}}[\mathcal{E}_{\omega-1}])=0$ and} \\
   & i_{\omega-1} < i \leq N-1,\\
+1 & \text{if CRC$(\bm{\hat{u}}[\mathcal{E}_{\omega-1}]) \neq 0$ and} \\
& i_{\omega-1} < i \leq N-1, i \neq i^*_\omega,\\
-1 & \text{if CRC$(\bm{\hat{u}}[\mathcal{E}_{\omega-1}]) \neq 0$ and}\\
   & i = i^*_\omega,
\end{cases}
\end{equation}
and for $\bm{\hat{u}}[\mathcal{E}_\omega]$ as
\begin{equation}
\label{equ:NDSCF:u_generalized}
\hat{u}[\mathcal{E}_\omega]_{i} = \frac{1 - \sgn(L[\mathcal{E}_{\omega}]_{0,i})T[\mathcal{E}_{\omega}]_i}{2} \text{,}
\end{equation}
where CRC$\left(\bm{\hat{u}}[\mathcal{E}_{\omega-1}]\right) = 0$ indicates a successful CRC verification, and CRC$\left(\bm{\hat{u}}[\mathcal{E}_{\omega-1}]\right) \neq 0$ indicates a failed CRC verification for $\bm{\hat{u}}[\mathcal{E}_{\omega-1}]$.

As all the network layers enable back-propagation, $\beta$ is trained with stochastic gradient descent technique \cite{DeepLearning}. In this paper, the following objective loss function is used:
\begin{equation}
\label{equ:NDSCF:loss}
\lambda = \sum_{\omega=1}^{2} l_\omega,
\end{equation}
where
\begin{equation}
\label{equ:NDSCF:l_omega}
\begin{split}
l_\omega & = \begin{cases}
\left(\hat{u}[\mathcal{E}_\omega]_{t_\omega} - u_{t_\omega} \right)^2 & \text{if CRC$\left(\bm{\hat{u}}[\mathcal{E}_\omega]\right) \neq 0$, }  i^*_\omega \geq t_\omega,\\
0 & \text{otherwise,}\\
\end{cases} \\
& \approx \begin{cases}
\frac{1}{\left(1+\exp{\left(L[\mathcal{E}_\omega]_{0,t_\omega}\right)}\right)^{2}} & \text{if CRC$\left(\bm{\hat{u}}[\mathcal{E}_\omega]\right) \neq 0$, } i^*_\omega \geq t_\omega,\\
0 & \text{otherwise,}\\
\end{cases} \\
\end{split}
\end{equation}
and $t_\omega$ is the first erroneous bit position of $\bm{\hat{u}}[\mathcal{E}_\omega]$ with $u_{t_\omega}=0$ as all-zero codewords are used. It is worth mentioning that the loss function is designed to penalize the wrong estimation of $i^*_\omega$ from the $Q^*(\mathcal{E}_\omega)$ layers only, not to penalize the wrong estimation of the message word $\bm{\hat{u}}[\mathcal{E}_\omega]$ obtained from the SC layers.

\section{Experimental Results}
\label{sec:experiment}

In this section, the performance of the proposed NDSCF decoder is evaluated in terms of FER and the average number of decoding attempts. Throughout this section, $\mathcal{P}(256,128)$ which is concatenated with a $24$-bit CRC is considered for the evaluation, similar to the case in Fig.~\ref{fig:DSCF:FER}. The NDSCF decoders are trained with $2.5\times10^5$ samples of $\bm{L}_n$ at each $E_b/N_0$ value, where $E_b/N_0 \in \{2,3,4,5\}$~dB. The parameter $\beta$ at each error order $\omega$ is initialized with a uniform distribution within the interval of $(0,10)$. The number of training epochs, mini-batch size, and learning rate are set to $50$, $256$, and $0.001$, respectively. During the evaluation phase, each decoder is tested with at least $10^5$ randomly generated codewords at each $E_b/N_0$ value, until at least $50$ frames in error are captured.

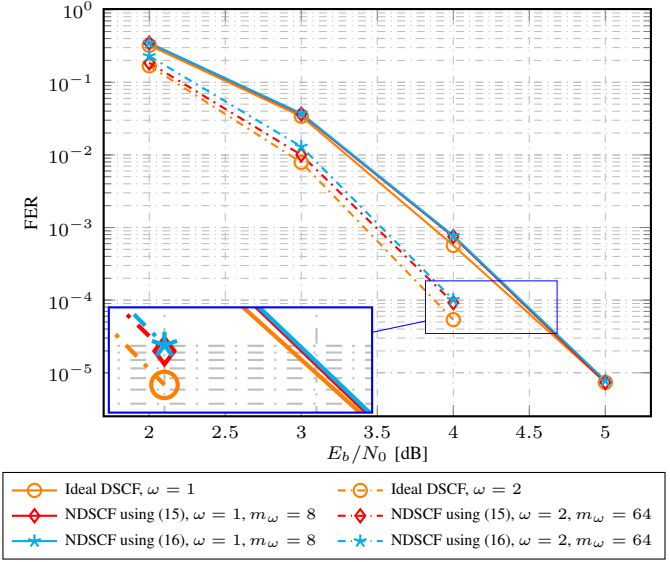
\begin{figure}[t]
	\centering
	\begin{tikzpicture}[spy using outlines = {rectangle, magnification=2.0, connect spies}]
\pgfplotsset{	
	label style = {font=\fontsize{7pt}{7}\selectfont},
	tick label style = {font=\fontsize{7pt}{7}\selectfont}
}

\begin{axis}[
scale = 1,
ymode=log,
xlabel={$E_b/N_0$ [\text{dB}]}, xlabel style={yshift=0.8em},
ylabel={FER}, ylabel style={yshift=-0.75em},
xtick={1,1.5,2,2.5,3,3.5,4,4.5,5},
ytick={1e-5,1e-4,1e-3,1e-2,1e-1,1e0},
grid=both,
ymajorgrids=true,
xmajorgrids=true,
grid style=dashdotted,
width=1\columnwidth, height=7cm,
thick,
mark size=2.5,
legend cell align={left},
legend style={
	at={(0,1e-5)},
	anchor=south west,
	column sep= 2mm,
	font=\fontsize{6pt}{7.2}\selectfont,
},
  legend to name=perf-legend-NDSCF,
  legend columns=2,
]

\addplot[
color=orange,
solid,
mark=o,
thick,
mark size=2.5,
]
table {
	2	0.322
	3	0.034
	4	0.000567568
	5	7.38E-06
};
\addlegendentry{\hspace*{-5pt}Ideal DSCF, $\omega=1$}

\addplot[
color=orange,
dashdotted,
mark=o,
mark options={solid},
thick,
mark size=2.5,
]
table {
	2	0.1683
	3	0.0079
	4	5.39E-05
};
\addlegendentry{\hspace*{-5pt}Ideal DSCF, $\omega=2$}

\addplot[
color=red,
solid,
mark=diamond,
thick,
mark size=2.5,
]
table {
	2	0.3462
	3	0.0365
	4	0.00075
	5	7.38E-06	
};
\addlegendentry{\hspace*{-5pt}NDSCF using (\ref{equ:NDSCF:NDSCF_LL}), $\omega=1$, $m_\omega=8$}

\addplot[
color=red,
dashdotted,
mark=diamond,
mark options={solid},
thick,
mark size=2.5,
]
table {
	2	0.185
	3	0.01
	4	9.22669E-05
};
\addlegendentry{\hspace*{-5pt}NDSCF using (\ref{equ:NDSCF:NDSCF_LL}), $\omega=2$, $m_\omega=64$}

\addplot[
color=cyan,
solid,
mark=star,
thick,
mark size=2.5,
]
table {
2	0.3475
3	0.0373
4	0.000772727
5	7.80E-06
};
\addlegendentry{\hspace*{-5pt}NDSCF using (\ref{equ:NDSCF:NDSCF_HW_LL}), $\omega=1$, $m_\omega=8$}

\addplot[
color=cyan,
dashdotted,
mark=star,
mark options={solid},
thick,
mark size=2.5,
]
table {
2	0.2295
3	0.013
4	0.000101416
};
\addlegendentry{\hspace*{-5pt}NDSCF using (\ref{equ:NDSCF:NDSCF_HW_LL}), $\omega=2$, $m_\omega=64$}

\coordinate (spypoint1) at (axis cs:4.25,0.8e-4);
\coordinate (magnifyglass1) at (axis cs:2.6,1.5e-5);

\coordinate (spypoint2) at (axis cs:4,0.7e-3);
\coordinate (magnifyglass2) at (axis cs:4.8,7e-2);
\end{axis}
\spy [blue, width=3.5cm, height=1.4cm] on (spypoint1) in node[fill=white] at (magnifyglass1);

\end{tikzpicture}
	\ref{perf-legend-NDSCF}
	\caption{FER of the proposed NDSCF decoding for $\mathcal{P}(256,128)$, concatenated with a $24$-bit CRC. The ideal DSCF decoder is also plotted as a reference.}
	\label{fig:NDSCF:FER}
\end{figure}

\begin{table}[t]
	\centering
	\caption{Optimized parameter $\beta$ of the proposed NDSCF decoders.}
	\setlength{\extrarowheight}{2.5pt}
	\begin{tabular}{lcccc}
		\toprule
		Metric & \multicolumn{2}{c}{(\ref{equ:NDSCF:NDSCF_LL})} & \multicolumn{2}{c}{(\ref{equ:NDSCF:NDSCF_HW_LL})} \\
		\cmidrule(lr){2-3}
		\cmidrule(lr){4-5}		
		$\omega$ & $1$ & $2$ & $1$ & $2$ \\
		\midrule
		$\beta$ & $2.206$ & $1.225$ & $2.801$ & $2.196$\\
		\bottomrule
	\end{tabular}
	\label{tab:param}
\end{table}

Fig.~\ref{fig:NDSCF:FER} illustrates the FER of the proposed NDSCF decoder for $\omega \in \{1,2\}$ and $m_\omega \in \{8,64\}$. The FER of the ideal DSCF decoder where the erroneous bits up to the $\omega$-th error order can always be accurately corrected is also plotted for comparison. Table~\ref{tab:param} specifies the optimized parameter $\beta$ for each bit-flipping metric in the proposed NDSCF decoder.

It can be observed from Fig.~\ref{fig:NDSCF:FER} that at a target FER of $10^{-4}$ and with $\omega=1$, the proposed NDSCF decoders have almost the same error-correction performance compared to that of the corresponding ideal DSCF decoder. With $\omega=2$ and at the target FER of $10^{-4}$, the proposed NDSCF decoders suffer from around $0.1$~dB of error-correction performance loss in comparison with the corresponding ideal DSCF decoder. Nevertheless, it can be seen that the introduction of the hardware-friendly metric in (\ref{equ:NDSCF:NDSCF_HW_LL}) incurs almost no error-correction performance loss in comparison with using the metric in (\ref{equ:NDSCF:NDSCF_LL}).

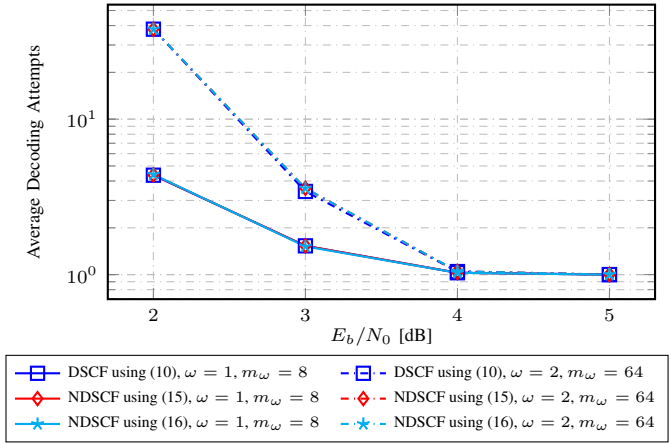
\begin{figure}[t]
	\centering
	\begin{tikzpicture}[spy using outlines = {rectangle, magnification=2, connect spies}]
\pgfplotsset{	
	label style = {font=\fontsize{7pt}{7}\selectfont},
	tick label style = {font=\fontsize{7pt}{7}\selectfont}
}

\begin{axis}[
scale = 1,
ymode=log,
xlabel={$E_b/N_0$ [\text{dB}]}, xlabel style={yshift=0.8em},
ylabel={Average Decoding Attempts}, ylabel style={yshift=-0.75em},
xtick={1,2,3,4,5},
grid=both,
ymajorgrids=true,
xmajorgrids=true,
grid style=dashdotted,
width=1\columnwidth, height=5.5cm,
thick,
mark size=2.5,
legend cell align={left},
legend style={
	at={(1,1)},
	anchor=north east,
	column sep= 2mm,
	font=\fontsize{6pt}{7.2}\selectfont,
},
  legend to name=latency-legend-NDSCF,
  legend columns=2,
]

\addplot[
color=blue,
solid,
mark=square,
thick,
mark size=2.5,
]
table {
2	4.3601
3	1.5316
4	1.028
5	1.000753358
};
\addlegendentry{\hspace*{-5pt}DSCF using (\ref{equ:DSCF:DSCF_LL}), $\omega=1$, $m_\omega=8$}

\addplot[
color=blue,
dashdotted,
mark=square,
mark options={solid},
thick,
mark size=2.5,
]
table {
2	37.9194
3	3.4292
4	1.046457364
5	1.0001
};
\addlegendentry{\hspace*{-5pt}DSCF using (\ref{equ:DSCF:DSCF_LL}), $\omega=2$, $m_\omega=64$}

\addplot[
color=red,
solid,
mark=diamond,
thick,
mark size=2.5,
]
table {
2	4.3763
3	1.5316
4	1.027758621
5	1.000771509
};
\addlegendentry{\hspace*{-5pt}NDSCF using (\ref{equ:NDSCF:NDSCF_LL}), $\omega=1$, $m_\omega=8$}

\addplot[
color=red,
dashdotted,
mark=diamond,
mark options={solid},
thick,
mark size=2.5,
]
table {
2	37.89
3	3.5873
4	1.05E+00
5	1.0001
};
\addlegendentry{\hspace*{-5pt}NDSCF using (\ref{equ:NDSCF:NDSCF_LL}), $\omega=2$, $m_\omega=64$}

\addplot[
color=cyan,
solid,
mark=star,
thick,
mark size=2.5,
]
table {
	2	4.4062
	3	1.518
	4	1.02619697
	5	1.0001
};
\addlegendentry{\hspace*{-5pt}NDSCF using (\ref{equ:NDSCF:NDSCF_HW_LL}), $\omega=1$, $m_\omega=8$}

\addplot[
color=cyan,
dashdotted,
mark=star,
mark options={solid},
thick,
mark size=2.5,
]
table {
	2	37.89
	3	3.5873
	4	1.05E+00
	5	1.0001
};
\addlegendentry{\hspace*{-5pt}NDSCF using (\ref{equ:NDSCF:NDSCF_HW_LL}), $\omega=2$, $m_\omega=64$}

\end{axis}

\end{tikzpicture}
	\ref{latency-legend-NDSCF}
	\caption{Average number of decoding attempts of the proposed NDSCF decoders in comparison with that of the DSCF decoder of \cite{DSCF} for $\mathcal{P}(256,128)$.}
	\label{fig:NDSCF:Latency}
	\vspace*{-0.5\baselineskip}
\end{figure}

Fig.~\ref{fig:NDSCF:Latency} depicts the average number of decoding attempts for the DSCF in \cite{DSCF} and the proposed NDSCF decoders. It can be seen that the average number of decoding attempts of the proposed NDSCF decoder is similar to that of the DSCF decoder under the same decoding configurations. Note that the average number of decoding attempts of all the decoders depicted in Fig.~\ref{fig:NDSCF:Latency} approaches $1$ at high $E_b/N_0$ values. This indicates that at high $E_b/N_0$ values, the complexity of the decoders approaches the complexity of a single SC decoder.

\section{Conclusion}
\label{sec:conclude}

In this paper, we proposed a neural dynamic successive cancellation flip (NDSCF) decoding algorithm of polar codes. The proposed decoder uses an additive parameter to improve the accuracy of the bit-flipping metric and the parameter is optimized by a novel training framework. The proposed decoder has the following advantages: (i) its decoding complexity approaches that of the successive cancellation (SC) decoding at high signal-to-noise ratio (SNR) regimes; (ii) only additions and multiplexers are needed during the course of decoding; (iii) negligible error correction performance loss is incurred in comparison with the ideal dynamic successive cancellation flip (DSCF) decoder; and (iv) it enables channel adaptation and simplifies over-the-air training as no labeling task is required and only all-zero codewords are needed during the training process. With the aforementioned advantages, NDSCF decoding is a potential candidate for practical applications in the 5G standard.

\section{Acknowledgment}
The authors would like to thank Arash Ardakani and Adam Cavatassi of McGill University for their helpful and constructive comments.
S. A. Hashemi is supported by a Postdoctoral Fellowship from the Natural Sciences and Engineering Research Council of Canada (NSERC).



\end{document}